\documentclass[aps,prl,twocolumn,showpacs,floatfix,superscriptaddress]{revtex4}
\usepackage{graphicx}
\usepackage{dcolumn}
\usepackage{amsmath}
\usepackage{amssymb}
\usepackage{dsfont}
\usepackage{bm}

\begin{document}

\title{Recurrence and P\'olya number of quantum walks}

\author{M. \v Stefa\v n\'ak\email[correspondence
  to:]{martin.stefanak@uni-ulm.de}} \affiliation{Department of
  Physics, FJFI \v CVUT, B\v rehov\'a 7, 115 19 Praha 1 - Star\'e
  M\v{e}sto, Czech Republic} \affiliation{Institute for Quantum
  Physics, University of Ulm, D-89069 Ulm, Germany}

\author{I. Jex} \affiliation{Department of Physics, FJFI \v CVUT, B\v
  rehov\'a 7, 115 19 Praha 1 - Star\'e M\v{e}sto, Czech Republic}

\author{T. Kiss} \affiliation{Department of Nonlinear and Quantum
  Optics, Research Institute for Solid State Physics and Optics,
  Hungarian Academy of Sciences, Konkoly-Thege u.29-33, H-1121
  Budapest, Hungary}


\date{\today}

\begin{abstract}
  We analyze the recurrence probability (P\'olya number) for
  $d$-dimensional unbiased quantum walks. A sufficient condition for a
  quantum walk to be recurrent is derived. As a by-product we find a
  simple criterion for localisation of quantum walks. In contrast to
  classical walks, where the P\'olya number is characteristic for the
  given dimension, the recurrence probability of a quantum walk
  depends in general on the topology of the walk, choice of the coin
  and the initial state. This allows to change the character of the
  quantum walk from recurrent to transient by altering the initial
  state.
\end{abstract}

\maketitle

Over the years random walks (RW) proved to be a useful tool in many
areas in physics as well as in other branches of science
\cite{overview}. Quantum analogues of random walks have been proposed
by Aharonov \cite{aharonov}. The quantum walk (QW) found a promising
application in quantum information for the construction of fast search
algorithms \cite{kempe}, which initiated considerable effort to
understand all aspects of QWs \cite{Grimmett}.  Recently, localisation
has been found in 2-D \cite{2dwBrun,2dw1,localisation} and in 1-D for
a generalized QW \cite{1dloc}.

Recurrence in the dynamics of physical systems is an important
phenomenon with many far reaching consequences
\cite{chandrasekhar}. In the strict sense it means the periodicity of
the system and in a broader sense it is the recurrence of some
particular property \cite{peres}. Recurrence phenomena have been
studied extensively for a variety of quantum systems
\cite{robinett}. In accordance with the terminology for the classical
random walks we will consider the recurrence probability that the
walker returns to the origin in QWs.

In the present paper we extend the concept of P\'olya number
characterizing the recurrence of RW to the quantum domain. We point
out the fundamental difference between the recurrence behaviour of
classical and quantum random walks. In particular, the recurrence of
QWs is not solely determined by the dimensionality of the walk, but
may depend on the topology of the walk, choice of the coin governing
the time evolution and the initial coin state. Moreover, due to the
intimate connection between localisation and recurrence, we find a
simple criterion for the former in QWs.

Let us consider an unbiased random walk on an infinite $d$-dimensional
lattice starting localized at the origin $\mathbf{0}$. The probability
that the walker returns to the origin during the time evolution is
called the P\'olya number of the walk. If the P\'olya number equals
one, then the walk is called {\it recurrent}, otherwise there is a
non-zero probability that the walker never returns to its starting
point. Such walks are {\it transient}. P\'olya \cite{polya} proved
that one and two-dimensional walks are recurrent while for higher
dimensions the RWs are transient and a unique P\'olya number is
associated to them in each dimension \cite{hughes}.

The P\'olya number of a classical random walk can be expressed with the probability $p_o(t)$ that the walker returns to the
origin in the $t$-th step and its behaviour depends solely on the infinite sum
$\sum_{t=0}^{\infty}p_o(t)$. A simple criterion exists for a RW to be transient: the series must converge, otherwise the walk is recurrent \cite{revesz}. The sequence $p_o(t)$ consists of positive numbers, therefore its decay determines the convergence of the series, if, e.g., it decays faster than $t^{-1}$ it is convergent.

The definition of the P\'olya number can be consistently extended for QWs  by the expression
\begin{equation}
P = 1-\prod\limits_{t=1}^{+\infty}(1-p_o(t)).
\label{polya:def}
\end{equation}
for an ensemble of identically prepared systems, in the following sense. Take a system and measure the position of the walker after one time step at the origin,
then discard the system. Take a second, identically prepared system and let it evolve for two time steps,
measure at the origin, then discard the system. Continue similarly for arbitrarily long evolution time.
The probability that the walker is found at the origin in a single series of such measurement records is the P\'olya number (\ref{polya:def}).
The above definition ensures that the same criterion applies as in the classical case: if
the series $\sum_{t=0}^{\infty}p_o(t)$ is convergent, then the walk is transient, otherwise it is recurrent \cite{jarnik}. Since any interaction with a quantum system and especially a measurement unavoidably disturbs it, it is important to fix how measurements are performed on quantum systems when comparing to classical results (cf. definitions for hitting time in graph QWs \cite{krovi,Kempe}).
One can imagine various experimental situations where quantum walks are observed during propagation.
The definition of recurrence should then be different to reflect properly the physics behind.

Let us briefly review unbiased QWs on $\mathbb{Z}^d$. The Hilbert space of the QW is given by the tensor product $\mathcal{H}=\mathcal{H}_P\otimes\mathcal{H}_C$ of the position space $\mathcal{H}_P=\ell^2(\mathds{Z}^d)$ with the basis formed by the states $|\mathbf{m}\rangle$ where $\mathbf{m}\in\mathbb{Z}^d$ and of the $c$-dimensional coin space $\mathcal{H}_C$. Here $c$ is determined by the topology of the walk, in particular by the number of neighbours to which the walker can jump in a single step. For example, the walker can jump in positive or negative direction in a single spatial dimension \cite{Grimmett}, which leads to $c=2d$. Alternatively, it can perform such jumps in all spatial dimension simultaneously \cite{2dqw}, so the coin space has the dimension $c=2^d$.
We denote all displacements possible in a single step of the walk by the set of shift vectors $\left\{\mathbf{e}_i|i=1,\ldots ,c\right\}$ and define the orthonormal basis in $\mathcal{H}_C$ formed by the states $|\mathbf{e}_i\rangle$. We consider unbiased walks where the shift vectors are restricted by the condition $\sum\mathbf{e}_i=0$. The time evolution operator which propagates a QW by a single step reads
\begin{equation}
U=S\cdot\left(I_P\otimes C\right),
\label{qw:time}
\end{equation}
where $I_P$ denotes the unit operator on the position space $\mathcal{H}_P$, $C$ is the coin flip operator acting on the coin state and the conditional step operator $S$ is defined by
\begin{equation}
S = \sum\limits_{i}|\mathbf{m}+\mathbf{e}_i\rangle\langle\mathbf{m}|\otimes|\mathbf{e}_i\rangle\langle\mathbf{e}_i|.
\end{equation}
The state of the QW after $t$ steps
\begin{equation}
|\psi(t)\rangle\equiv\sum\limits_{\mathbf{m},i}\psi_i(\mathbf{m},t)|\mathbf{m}\rangle\otimes|\mathbf{e}_i\rangle=U^t|\psi(0)\rangle
\label{time:evol}
\end{equation}
is given by successive application of the operator (\ref{qw:time}) on the initial state $|\psi(0)\rangle$. For the above defined walk to be unbiased the coin flip $C$ must be a unitary operator acting on $\mathcal{H}_C$ with all matrix elements $C_{ij}\equiv\langle\mathbf{e}_i|C|\mathbf{e}_j\rangle$ of the same absolute value $1/\sqrt{c}$.
Such matrices are related to the Hadamard matrices \cite{dita}.

Due to the translational invariance of the QWs in consideration, the time evolution (\ref{time:evol}) is greatly simplified with the help of the Fourier transformation
\begin{equation}
\tilde{\psi}(\mathbf{k},t)\equiv\sum\limits_\mathbf{m}\psi(\mathbf{m},t) e^{i (\mathbf{m}\cdot\mathbf{k})}.
\label{qw:ft}
\end{equation}
Here we have defined the $c$-component vectors
\begin{equation}
\psi(\mathbf{m},t)\equiv{\left(\psi_1(\mathbf{m},t),\psi_2(\mathbf{m},t),\ldots,\psi_c(\mathbf{m},t)\right)}^T.
\label{prob:ampl}
\end{equation}
The Fourier transformation (\ref{qw:ft}) defines an isometry between $\ell^2(\mathds{Z}^d)$ and $L^2(\mathds{K}^d)$ where $\mathds{K}=(-\pi,\pi]$ can be thought of as the phase of a unit circle in $\mathds{R}^2$ and $\mathbf{k}\in\mathbb{K}^d$. The time evolution in the Fourier picture simplifies into
\begin{equation}
\tilde{\psi}(\mathbf{k},t)=\widetilde{U}(\mathbf{k})\tilde{\psi}(\mathbf{k},t-1),\quad \widetilde{U}(\mathbf{k})\equiv D(\mathbf{k})\cdot C,
\label{qw:te:fourier}
\end{equation}
where the time evolution operator in the Fourier picture $\widetilde{U}(\mathbf{k})$ is given by the coin $C$ and the $c$-dimensional diagonal matrix $D(\mathbf{k})$ with $j$th diagonal element $e^{-i\mathbf{e}_j\cdot\mathbf{k}}$,
determined by the topology of the QW. The time evolution in the Fourier picture (\ref{qw:te:fourier}) is solved by diagonalising the matrix $\widetilde{U}(\mathbf{k})$ which is possible due to the fact that by construction this matrix is unitary. Hence it has eigenvalues of the form
$\lambda_j(\mathbf{k})=\exp{\left(i \omega_j(\mathbf{k})\right)}$. We denote the corresponding eigenvectors as $v_j(\mathbf{k})$. With this notation we can write the state of the walker in the Fourier picture at time $t$ in the form
\begin{equation}
\tilde{\psi}(\mathbf{k},t) = \sum_j e^{i\omega_j(\mathbf{k})t}\left(\tilde{\psi}(\mathbf{k},0),v_j(\mathbf{k})\right)v_j(\mathbf{k}).
\label{sol:k}
\end{equation}
Performing the inverse Fourier transformation we can obtain the probability amplitudes $\psi(\mathbf{m},t)$. As we are interested in the recurrence of the QWs we need the probability that the walker returns to the origin at time $t$
\begin{equation}
p_o(t)\equiv p(\mathbf{0},t)=\left\|\psi(\mathbf{0},t)\right\|^2.
\label{po}
\end{equation}
Moreover, in analogy with the classical problem, we consider initial states which are localized at the origin, i.e. $\psi(\mathbf{m},0)=\mathbf{0}$ for all $\mathbf{m}\neq\mathbf{0}$. From the definition (\ref{qw:ft}) follows that under such restrictions the Fourier transformation of the initial condition $\tilde{\psi}(\mathbf{k},0)$ entering (\ref{sol:k}) is constant and equals $\psi(\mathbf{0},0)$. Hence we find the exact expression for the probability amplitude determining the recurrence behaviour of the QW
\begin{eqnarray}
\nonumber \psi(\mathbf{0},t) & = & \sum_{j=1}^c I_j(t),\quad I_j(t) =  \int\limits_{\mathbb{K}^d}\frac{d\mathbf{k}}{(2\pi)^d}e^{i\omega_j(\mathbf{k})t}\cdot f_j(\mathbf{k}),\\  f_j(\mathbf{k}) & = & \left(\psi(\mathbf{0},0),v_j(\mathbf{k})\right)\cdot v_j(\mathbf{k})\, .
\label{psi:0}
\end{eqnarray}

The recurrence behaviour of a classical random walk is uniquely determined by its dimensionality. In contrast, for quantum walks, as seen from (\ref{psi:0}), one has more freedom: both the initial state $\psi(\mathbf{0},0)$ and the coin, represented by the phases $\omega_j(\mathbf{k})$ and the eigenvectors $v_j(\mathbf{k})$, can be varied. Moreover, both $\omega_j(\mathbf{k})$ and $v_j(\mathbf{k})$ are affected by the topology of the QW through the matrix $D(\mathbf{k})$. In the following we demonstrate that the recurrence of QWs can be in fact altered by exploiting the additional freedom offered by quantum mechanics.

The recurrence of the QW is determined by the asymptotics of (\ref{psi:0}) which can be calculated e.g. by the method of stationary phase \cite{statphase}. Accordingly, the asymptotic behaviour of $I_j(t)$ is only affected by the points where all derivatives of $\omega_j(\mathbf{k})$ vanish (saddle points). In particular, it is determined by the degeneracy of the saddle points, given by the rank of the Hessian matrix, and the cardinality of the set of saddle points. The classical result \cite{statphase} shows that if the phase $\omega_j(\mathbf{k})$ has no saddle points then $I_j(t)$ decays exponentially. If $\omega_j(\mathbf{k})$ has finitely many saddle points which are non-degenerate, i.e. the second derivative of $\omega_j(\mathbf{k})$ is non-vanishing, then asymptotically $I_j(t)$ is given by $I_j(t)\sim t^{-d/2}$. Here we assume that $f_j(\mathbf{k})$ is smooth and non-vanishing at the saddle points. However, if the initial state $\psi(\mathbf{0},0)$ is orthogonal to some eigenvector $v_j(\mathbf{k})$ then $f_j(\mathbf{k})=0$ and the corresponding integral is zero. If this is true only for $\mathbf{k}$ corresponding to one of the saddle points then this particular saddle point will not contribute to the integral $I_j(t)$. Hence the choice of the initial state might change the behaviour of the QW from recurrent to transient. Moreover, we can encounter situations where the phase $\omega_j(\mathbf{k})$ does not depend explicitly on some of the variables $k_i$ or has a continuum of saddle points. In such a case we can expect a slow-down of the decay of $I_j(t)$.

Let us illustrate these features on several examples. We consider QWs for which the shift vectors $\mathbf{e}_i$ have all entries equal to $\pm 1$, i.e. the coin space has the dimension $c=2^d$. For this particular type of QWs we find that the diagonal matrix $D(\mathbf{k})$ can be written as a tensor product
\begin{equation}
D(\textbf{k}) = D(k_1)\otimes\ldots\otimes D(k_d)
\label{dk2}
\end{equation}
of $2\times 2$ diagonal matrices $D(k_j)=\textrm{diag}(e^{-ik_j},e^{ik_j})$.

First, we present examples where the phases $\omega_j(\mathbf{k})$ have finitely many non-degenerate saddle points. As follows from the above discussion the probability $p_o(t)$ behaves asymptotically like $t^{-d}$ where $d$ is the dimension of the walk. We start with general unbiased 1-D QW
\begin{equation}
C(\alpha,\beta)=\frac{1}{\sqrt{2}}\left(
                      \begin{array}{rr}
                        e^{i\alpha} & e^{-i\beta} \\
                        e^{i\beta} & -e^{-i\alpha} \\
                      \end{array}
                    \right).
\label{1D}
\end{equation}
Here the matrix $\widetilde{U}(\alpha,\beta,k)\equiv D(k)\cdot C(\alpha,\beta)$ has eigenvalues $\lambda_j(\alpha,k)=\pm e^{\pm i\omega(\alpha,k)}$ with the phase
\begin{equation}
\sin\omega(\alpha,k)=-\frac{\sin{(k-\alpha)}}{\sqrt{2}}.
\label{angle:had}
\end{equation}
We find that the phase $\omega(\alpha,k)$ has saddle points $k^0=\alpha\pm\pi/2$ and hence $p_o(t)$ behaves asymptotically like $t^{-1}$. Moreover, the asymptotic behaviour is independent of the initial state. Indeed, no non-zero initial state $\psi(\mathbf{0},0)$ exists which is orthogonal to both eigenvectors for $k^0=\alpha\pm\pi/2$. Comparing the decay $p_o(t)\sim t^{-1}$ with the recurrence criterion \cite{revesz} we find that all unbiased 1-D QWs are recurrent for all initial states in concord with the classical result. However, none of the QWs from the class (\ref{1D}) exhibits localisation, since for all of them the probability $p_o(t)$ converges to zero. One can achieve localisation in 1-D by considering generalized QWs for which the coin has more degrees of freedom \cite{1dloc,prep}.

Let us now turn to 2-D QWs for which the coin factorizes into a tensor product
\begin{equation}
C(\bm \alpha,\bm \beta)\equiv C(\alpha_1,\beta_1)\otimes C(\alpha_2,\beta_2),
\label{H2d}
\end{equation}
i.e. we have an independent coin for each spatial dimension. From the relation (\ref{dk2}) we find that the matrix $\widetilde{U}(k_1,k_2)$ also has the form of a tensor product
\begin{equation}
\widetilde{U}(k_1,k_2)=\widetilde{U}_1(k_1)\otimes\widetilde{U}_2(k_2).
\label{H2}
\end{equation}
Therefore the eigenvalues $\lambda_j(\mathbf{k})$ of (\ref{H2}) factorize into the eigenvalues of $\widetilde{U}_1$ and $\widetilde{U}_2$. Hence their phases $\omega_j(\mathbf{k})$ have non-degenerate saddle points and we find that $p_o(t)$ behaves asymptotically like $t^{-2}$. Similarly to the 1-D case this scaling is identical for all initial coin states. Hence we find that all unbiased 2-D QWs driven by independent coins for each spatial dimension (\ref{H2d}) are transient for all initial states. The P\'olya number of this class of QWs is approximately 0.29143. We note that the concept of QWs with independent coins can be extended to arbitrary dimension $d$ and their P\'olya numbers are uniquely determined by $d$ \cite{prep}. In contrast with the classical RWs they are recurrent only for $d=1$. Nevertheless, the transience of $d$-dimensional QWs with independent coins for $d>1$ is not surprising when we recall that the QWs spread quadratically faster compared to the classical RWs. However, the probability $p_o(t)$, which determines the recurrence of the walk, does not need to decay faster for the QW compared to the classical RW. In fact, the constructive interference can be so strong that the QW will exhibit localisation \cite{2dw1,localisation,1dloc}.

Let us now analyze the emergence of localisation, or more generally the slow-down of the decay of $I_j(t)$. Consider the situation where $\omega_j(\mathbf{k})$ does not depend on some of the variables $k_i$, say $n$ of them. This opens up the possibility that $I_j(t)$ in (\ref{psi:0}) factorizes into the product of time-independent and time-dependent integrals over $n$ and $d-n$ variables. If in the reduced space of $d-n$ dimensions a finite number of non-degenerate saddle points are found, then one can proceed similarly to the previous case and find the asymptotic behaviour
\begin{equation}
p_o(t)\sim t^{-(d-n)}.
\label{asymp:n}
\end{equation}
Comparing this expression with the recurrence condition \cite{revesz} of random walks we find a sufficient condition for the recurrence of QWs: at least one eigenvalue of the matrix $C(\mathbf{k})$ should be such that its phase $\omega_j(\mathbf{k})$ depends explicitly at most on a single variable and has a finite number of saddle points in the reduced space. Moreover, if $\omega_j(\mathbf{k})$ is independent of $\mathbf{k}$ the leading order term of the probability $p_o(t)$ is a constant. In the latter case $p_o(t)$ has a non-vanishing limit value and therefore the QW exhibits localisation for initial states being non-orthogonal to the corresponding eigenvector. However, as we have already pointed out, a given initial state $\psi(\mathbf{0},0)$ might lead to a faster decay than (\ref{asymp:n}) if it is orthogonal to all eigenvectors corresponding to such eigenvalues. Orthogonality must hold at least at all the related saddle points.

The above discussed situation is nicely illustrated on the example of the 2-D Grover walk driven by the coin $G$ with the matrix elements $G_{m,n}=1/2-\delta_{mn}$. This QW has been extensively studied \cite{2dqw,2dw1,localisation}. It was first identified numerically \cite{2dw1} and later proven analytically \cite{localisation} by considering the degeneracy of the eigenvalues of the transfer matrix that the Grover walk exhibits localisation except for a particular initial state
\begin{equation}
\psi(\mathbf{0},0) = \psi_G\equiv \frac{1}{2}\left(1,-1,-1,1\right)^T.
\label{grover:nospike:state}
\end{equation}
For the 2-D Grover walk the matrix $\widetilde{U}(\mathbf{k})$ has two constant eigenvalues $\lambda_{1,2}=\pm 1$ and two of the form
\begin{equation}
\label{eigenval:Grover}
\lambda_{3,4}(\mathbf{k}) = e^{\pm i \omega(\mathbf{k})},\ \cos(\omega(\mathbf{k})) = -\cos{k_1}\cos{k_2}.
\end{equation}
Hence the 2-D Grover walk exhibits localisation unless the initial state is orthogonal to the eigenvectors corresponding to $\lambda_{1,2}$ at every point $\mathbf{k}=(k_1,k_2)$. The analysis of the eigenvectors of the matrix $\widetilde{U}(\mathbf{k})$ reveals that such a vector is unique and equals (\ref{grover:nospike:state}). Moreover, for the particular initial state (\ref{grover:nospike:state}) the probability $p_o(t)$ decays like $t^{-2}$, since the asymptotic behaviour is determined by the remaining eigenvalues $\lambda_{3,4}(\mathbf{k})$ for which the phase $\omega(\mathbf{k})$ has only non-degenerate saddle points. Therefore the 2-D Grover walk is recurrent for all initial states except the one given in (\ref{grover:nospike:state}). The numerical value of P\'olya number for the particular initial state (\ref{grover:nospike:state}) equals to that of the 2-D QWs with independent coins \cite{prep}. As a generalization of the Grover walk, one can construct for arbitrary dimensions a QW which is recurrent \cite{prep}, except for a subspace of initial coin states. Moreover, this QW exhibits localisation in even dimensions.

Finally, we consider the situation where the phase $\omega_j(\mathbf{k})$ has a continuum of saddle points, e.g. they align on some curve
$\gamma$. The previously discussed case of $\omega_j(\mathbf{k})$ which does not depend on $n$ variables can be considered as a particular example of this more general situation, since such $\omega_j(\mathbf{k})$ obviously has a zero derivative with respect to those $n$ variables. The case of 2-D integrals with curves of stationary points are treated in \cite{statphase} where it is shown that the continuum of stationary points slows-down the decay of such integrals to $t^{-1/2}$. Similar results can be expected for higher dimensional saddle domains, however, much less is known about the stationary phase method for that case.

The recurrence of the 2-D Fourier walk can be analyzed along these lines. It is driven by the coin $F$ with the matrix elements
\begin{equation}
F_{m,n}=\frac{1}{2} \exp{\left[i\pi (m-1)(n-1)/2\right]}.
\end{equation}
Here we find that all four phases $\omega_j(\mathbf{k})$ of the time evolution operator in the Fourier picture $\widetilde{U}(\mathbf{k})$ have saddle points at $k_1^0=\pi/4,-3\pi/4$ and $k_2^0=\pm\pi/2$ and $\omega_{1,2}(\mathbf{k})$ have two saddle lines $\gamma_1=(k_1,0)$ and $\gamma_2=(k_1,\pi)$ \cite{prep}. Hence $p_o(t)$ behaves asymptotically like $t^{-1}$ and the 2-D Fourier walk is recurrent, except for the subspace
\begin{equation}
\psi_F(a,b) = \left(a,b,a,-b\right)^T,\quad 2|a|^2+2|b|^2=1,
\label{psi:F}
\end{equation}
of the initial states which are orthogonal to the eigenvectors $v_{1,2}(\mathbf{k})$ for $\mathbf{k}\in\gamma_{1,2}$. For the initial states of the family (\ref{psi:F}) we find the asymptotic behaviour $p_o(t)\sim t^{-2}$ implying that the walk is transient. This corresponds to the absence of the central spikes in the spatial distribution, as it was found numerically for a special case in \cite{2dw1}.
The value of the P\'olya number for the family (\ref{psi:F}) can be varied by the complex parameters $a$ and $b$ of the initial state \cite{prep}.

To conclude, we have extended the concept of P\'olya number to QWs in order to characterize their recurrence properties. Our main result is that, unlike in the classical case, recurrence for QWs can depend not only on the dimensionality of the lattice but also on the topology of the walk, choice of the coin operator and the initial state. We have formulated sufficient conditions for the recurrence and localisation of QWs. The present study of the recurrence and P\'olya number based on the properties of the matrix $\widetilde{U}(\mathbf{k})$ can be extended to higher dimensional QWs where additional interesting effects can be expected \cite{prep}. The explicit dependence of the QW on the coin and the initial coin state opens up the possibility to design the value of the P\'olya number.

M. \v S. thanks W. P. Schleich for stimulating discussions. The financial support by MSM 6840770039, M\v SMT LC 06002, the Czech-Hungarian cooperation project (KONTAKT,CZ-2/2005), by the Hungarian Scientific Research Fund (T049234) and EU project QUELE is gratefully acknowledged. Moreover, M. \v S. is grateful for financial support from the EU in the framework of CONQUEST.

\end{document}